\documentclass[a4paper,twocolumn,showpacs,amsmath,pra,amssymb,nofootinbib]{revtex4-1} %letterpaper,preprint 
\usepackage[T1]{fontenc}
\usepackage[latin9]{inputenc} 
\usepackage{times}
\usepackage{color}
\usepackage{xspace}
\usepackage{amssymb,amsmath}
\usepackage{amsbsy}
\usepackage{graphicx}
\usepackage{epstopdf}
\usepackage{float}

\newcommand{\diff}{\mathrm{d}}

\newcommand{\br}{\ensuremath{\mathbf{r}}\xspace}

\newcommand{\kt}{\ensuremath{\kb T}\xspace}
\newcommand{\kb}{\ensuremath{k_B}\xspace}

\newcommand{\bx}{\ensuremath{\mathbf{x}}\xspace}
\newcommand{\x}{\mathbf{x}}
\newcommand{\R}{\mathbf{R}}
\renewcommand{\r}{\mathbf{r}}
\newcommand{\xd}{\mathbf{x}^\prime}
\renewcommand{\k}{\mathbf{k}}
\newcommand{\kd}{\mathbf{k}^\prime}

\begin{document}
\newcommand{\bk}{\mathbf{k}}
\newcommand{\nt}{\tilde{n}}
\newcommand{\Lnh}{\ensuremath{\mathcal{L}}\xspace}
\newcommand{\dbx}{\diff\bx}
\newcommand{\dbk}{\diff\bk}
\newcommand{\nbe}{\ensuremath{\bar{n}_{\mathrm{BE}}}\xspace}

\title{Thermodynamics and coherence of a trapped dipolar Fermi gas}
\author{D.~Baillie} 
\author{P.~B.~Blakie}  

\affiliation{Jack Dodd Centre for Quantum Technology, Department of Physics, University of Otago, Dunedin, New Zealand.}

\begin{abstract}
 We develop a meanfield treatment of a polarized trapped  Fermi gas with dipole-dipole interactions. Our approach is based on self-consistent semiclassical Hartree-Fock theory that accounts for direct and exchange interactions. We discuss our procedure for numerically implementing the calculation. We study the thermodynamic  and  the first and second order correlation properties of the system. We find that the system entropy depends on the trap geometry, allowing the system to be cooled as the trap aspect ratio is increased, and that exchange interactions cause the correlation functions to be anisotropic in the low temperature regime. We also find that many uniform gas thermodynamic predictions,  for which direct interaction effects vanish, are qualitatively unreliable for  trapped systems, most notably for oblate traps. We develop a   simplified Hartree formalism that is applicable to anisotropic harmonic traps.
 
\end{abstract}

\pacs{03.75.Ss, 74.20.Rp, 67.30.H-,05.30.Fk} 

\maketitle

%==============================================================================
\section{Introduction}

There has been phenomenal recent progress towards producing a quantum degenerate Fermi gas of polar molecules \cite{Ni2008a,Ospelkaus2010a,Ni2010a}.  The long-range nature of the dipolar interaction and the  strong dipole moments of ground state heteronuclear molecules has opened the door to an exciting array of new physics \cite{Fregoso2009a,Baranov2008a,Baranov2008b,Pupillo2008a,Xin2008a,Ortner2009a,Shi2010a,Wu2010a}. Of particular interest is the superconducting regime of this system, first considered in Ref.~\cite{You1999a} and developed in \cite{Baranov2002a,Baranov2004a,Baranov2005a}.
  
Here we are concerned with a Hartree-Fock description of the trapped normal system in the semiclassical approximation. Such a formalism has been the workhorse theory for computing thermodynamic properties in ultra-cold gases with contact interactions, however it is considerably more difficult to evaluate in systems with long-range interactions. We begin by briefly surveying   previous work in this area.
 Initial work  for the zero temperature case   by G{\'o}ral \textit{et al.}~\cite{Goral2001b} used a Thomas-Fermi ansatz. 
In that approach an isotropic momentum distribution was assumed causing the exchange (Fock) term to vanish, nevertheless the retention  of the direct (Hartree) interaction was observed to distort the system  density distribution from the aspect ratio of the trapping potential. A variational treatment by Miyakawa \textit{et al.}~\cite{Miyakawa2008a} extended the Thomas-Fermi ansatz to allow for an ellipsoidal momentum distribution and showed that exchange effects tended to distort the momentum distribution of the system. Beyond the variational approximation, Zhang and Yi \cite{Zhang2009a} solved the full semiclassical theory for the case of a cylindrical trap by minimising the energy function of the system using a Monte-Carlo method. Lin \textit{et al.}~\cite{Lin2010a} have presented a similar formalism but in terms of a self-consistent calculation of the phase-space distribution (Wigner function). We also  note  analytical results by Chan \textit{et al.}~\cite{Chan2010a} developed for a Fermi-liquid description of the dipolar Fermi gas, and work in the zero-temperature regime  on dynamic properties such as collective modes and expansion dynamics \cite{Goral2003a,He2008a,Sogo2009a,Lima2010a}, and sound propagation \cite{Ronen2010a}. There has also been some studies using  time-dependent Hartree-Fock theory for this system \cite{Tohyama2009a,Lima2010b}.

For the finite temperature, regime Endo \textit{et al.}~\cite{Endo2010a} have developed a variational approach, formally valid in the non-degenerate regime, which allowed them to assess the effect of temperature on the position and momentum space distortions. Zhang and Yi \cite{Zhang2010a} have also presented results of a full semiclassical Hartree-Fock calculation, which they have used to explore the system deformation and stability.

In this paper we present the results of a fully self-consistent Hartree-Fock treatment of the trapped gas similar to the approach employed in Ref.~\cite{Zhang2010a}.  As was revealed in \cite{Zhang2009a}, for many system properties (particularly stability) this level of theory provides large corrections over the variational predictions. Furthermore, in the (finite temperature) degenerate regime $T\le T_F^0$, there are currently no valid variational theories. This regime is urgently in need of  quantitative theoretical analysis as  experiments with polar molecules are now approaching degeneracy.

The outline of the paper is as follows. After introducing the basic Hartree-Fock formalism we compare our results against others in the literature for the distortion of the system position and momentum distribution. We then calculate a wide range of thermodynamic quantities of the system, including the first calculations for  entropy and heat capacity of the trapped gas. Using these results we show that adiabatic mechanical deformation of the trapping potential can be used to change the temperature of the gas, in particular that compressing the trap in the polarization direction will cause the gas to reduce in temperature. We then discuss how to define and calculate correlation functions for the system. Our results show that exchange interaction effects cause the correlation functions to be anisotropic in the low temperature regime. We introduce a simplified Hartree theory in which exchange effects are neglected, and present results demonstrating the accuracy and applicability of this theory. We give a full account of our numerical method and the techniques we employ to make the calculations tractable and accurate in the Appendix.

\section{Theory}
\subsection{Formalism}
We consider a gas of spin polarized fermions that interact by a long-range dipole-dipole interaction of the form
\begin{equation}
U_{dd}(\x)=\frac{C_{dd}}{4\pi}\frac{1-3\cos^2\theta}{|\x|^3},\label{eqnUdd}
\end{equation}
where $C_{dd}=d^2/\epsilon_0$, with $d$ the electric dipole moment, and with $\theta$ the angle between the position vector and the polarization axis, which we take to be the $z$ direction. The atoms are confined within a cylindrically symmetric harmonic trap  
\begin{equation}
U(\x)=\frac{m}{2}\left[\omega_\rho^2(x^2+y^2)+\omega_z^2z^2\right],\label{eqn_trap}
\end{equation}
with aspect ratio $\lambda=\omega_z/\omega_\rho$.

In the semi-classical approach  the system at temperature $T$ is described by the Wigner function  
\begin{equation}
W(\x,\k)=\frac{1}{\exp([\epsilon(\x,\k)-\mu]/k_BT)+1},\label{eqnWigner}
\end{equation}
where $\mu$ is the chemical potential and the Hartree-Fock dispersion relation is
\begin{equation}
\epsilon(\x,\k)=\frac{\hbar^2k^2}{2m}+U(\x)+\Phi_D(\x)-\Phi_E(\x,\k),\label{eqn_epsilon}
\end{equation}
with
\begin{align}
\Phi_D(\x)&=\int \frac{d\xd d\kd}{(2\pi)^3}U_{dd}(\x-\xd)W(\xd,\kd),\label{PhiD}\\
\Phi_E(\x,\k)&=\int \frac{ d\kd}{(2\pi)^3}\tilde{U}_{dd}(\k-\kd)W(\x,\kd),\label{PhiE}
\end{align}
the direct and exchange interaction terms, respectively. In Eq.~(\ref{PhiE}) $\tilde{U}_{dd}$ is the Fourier transform of the dipole-dipole interaction, given by 
\begin{equation}
\tilde{U}_{dd}(\k)=\frac{C_{dd}}{3}[3\cos^2\theta_\k-1],\label{eqnUddk}
\end{equation}
 where $\theta_\k$ is the angle between $\k$ and $k_z$.
 
To find equilibrium solutions Eqs.~(\ref{eqnWigner})-(\ref{PhiE}) must be solved self-consistently subject to the additional constraint of atom number, i.e.
\begin{equation} N=\int \frac{d\x d\k}{(2\pi)^3} W(\x,\k),
\end{equation}
fixed by adjusting the chemical potential.

We note that the Hartree-Fock dispersion relation  (\ref{eqn_epsilon}) can be derived by minimizing the free energy (e.g.~see \cite{Bergeman1997a}) for which the total energy is given by \cite{Goral2001b} 
\begin{equation}
    E\hspace{-0.7mm}  =\hspace{-2mm}\int\hspace{-1mm} \frac{d\x d\k}{(2\pi)^3}\left[\frac{\hbar^2k^2}{2m}+U(\x)+\frac{\Phi_D(\x)}{2}-\frac{\Phi_E(\x,\k)}{2}\right]\hspace{-1mm}W(\x,\k).\label{Efunc}
\end{equation}

\subsection{Numerical treatment}
The semiclassical approach has seen extensive application to  ultra-cold Bose and Fermi gases with contact interactions, and has become the \textit{de facto} standard theory for providing thermodynamic information. 
The numerical solution of the semi-classical theory for the dipolar gas is rather more difficult, with the following main challenges: 
\begin{itemize}
\item[(i)] Interactions are non-local requiring careful choice of numerical grids and techniques to deal with the required convolutions.
\item[(ii)] The exchange interaction term  ($\Phi_E$) depends on both position and momentum variables.
\end{itemize}
In regard to point (i), extensive work on dipolar Bose and Fermi gases (e.g.~see \cite{Goral2001a,Ronen2006b,Blakie2009E}) has seen the development of  numerical tools for dealing with the direct potential ($\Phi_D$), such as the use of Hankel transforms for exploiting the cylindrical symmetry and allowing convolutions to be performed in Fourier space efficiently (e.g.~see \cite{Ronen2006a}).
However, point (ii) remains a considerable challenge and demands a self consistent solution of $W(\x,\k)$ in full $\x$- and $\k$-space   for the trapped case. In contrast, for the case of fermions or bosons with local  interactions $\Phi_E(\x,\k)\to\Phi_E(\x)$ and the momentum dependence of the excitations is only through the kinetic energy term. Since this momentum dependence is of a simple (isotropic) form it can be integrated out to provide a description only dependent on the density (see Sec.~\ref{SecHthry}).
 Thus, the full Wigner function does not need to be constructed to obtain a self-consistent solution. Furthermore, with local interactions the density is only a function of the local value of the external potential and the calculation can be mapped to a single scalar variable avoiding the need for a full spatial representation.

In the dipolar gas, symmetry is  broken by the polarization direction of the dipoles and in general the most symmetry the system can exhibit (even in an isotropic harmonic trap) is cylindrical.  Because the symmetry axis of the trap we consider (\ref{eqn_trap}) coincides with the polarization direction we can exploit this symmetry to simplify the Wigner function to a function four variables $\{\rho,z,k_\rho,k_z\}$, i.e.~ cylindrical coordinates in $\x$ and $\k$ space.

In implementing our numerical algorithm we make considerable use of Fourier transform techniques and quadrature based on discrete cosine and Bessel functions. Extensive details of the numerical algorithm are given in Appendix \ref{s:numericalmethods}.

\section{Results\label{s:results}}
In this section we present our results for the thermodynamic properties of a trapped dipolar Fermi gas. Following the standard convention (e.g.~see \cite{Zhang2010a}) we characterize our interaction strength in terms of dimensionless parameter
 $D_t=N^{1/6}C_{dd}/4\pi\hbar{\omega}{a}_{ho}^3$, where ${\omega}$ is the geometric mean trap frequency and $ {a}_{ho}=\sqrt{\hbar/m{\omega}}$.  Another energy scale we use is the ideal Fermi temperature $T_F^0=\hbar\omega(6N)^{1/3}/k_B$ for the harmonically trapped gas.  By parameterizing the interaction in terms of $D_t$, the temperature in units of $T_F^0$ and lengths in units of $N^{1/3}a_{ho}$, it is easy to show that the analysis is independent of $N$ (see \cite{Goral2001b}).
 
 The interplay between trapping geometry and interactions is of central importance in the dipolar gas.  To explore this relationship we present results for the cases of prolate ($\lambda=0.1$), spherical ($\lambda=1$) and oblate ($\lambda=10$) trapping. As the dipolar interaction has an attractive component, the system is not stable for all parameter regimes, as has been studied extensively (e.g.~see \cite{Goral2001b,Miyakawa2008a,Zhang2009a}). In general, oblate trapping allows the largest values of $D_t$ before the onset of collapse, as this geometry enhances the repulsive aspect of the long-range interactions. It has also been shown that thermal fluctuations tend to stabilize the system somewhat against collapse \cite{Zhang2010a}, so that the strictest condition on stability occurs at $T=0$. For values of $D_t\le1$, all three trapping geometries we consider are stable at $T=0$. For $D_t=2$ only the oblate configuration of $\lambda=10$ is stable at $T=0$, while the $\lambda=1$ and $\lambda=0.1$ cases become stable at about $0.3T_F^0-0.5T_F^0$.

\subsection{Thermal properties of a trapped dipolar Fermi gas}
\subsubsection{Position and momentum space distortion}
The anisotropy of the dipole-dipole interaction is reflected in the equilibrium distribution of the system. This is conveniently characterized by two simple parameters (see \cite{Zhang2010a})
\begin{align} 
\alpha&\equiv \sqrt{\frac{\langle k_x^2\rangle}{\langle k_z^2\rangle}},\\
\beta&\equiv\frac{1}{\lambda}\sqrt{\frac{\langle x^2\rangle}{\langle z^2\rangle}},
\end{align}
quantifying the momentum and position space distortion of the system, respectively, where
\begin{equation}
\langle x^2\rangle \equiv  \int \frac{d\x d\k}{(2\pi)^3} \,x^2\,W(\x,\k),
\end{equation}
is the $x$ variance, etc.
 The trap anisotropy appears in the definition of $\beta$ so that in the absence of interactions, where the position density has the  anisotropy imposed by the trap, we have $\beta=1$.  In contrast the non-interacting momentum distribution is spherically symmetric. 

\begin{figure}[!tbh]
\begin{center}
%load cutoff;runVariational;plotAlphaBetaDt
    \includegraphics[width=3.4in]{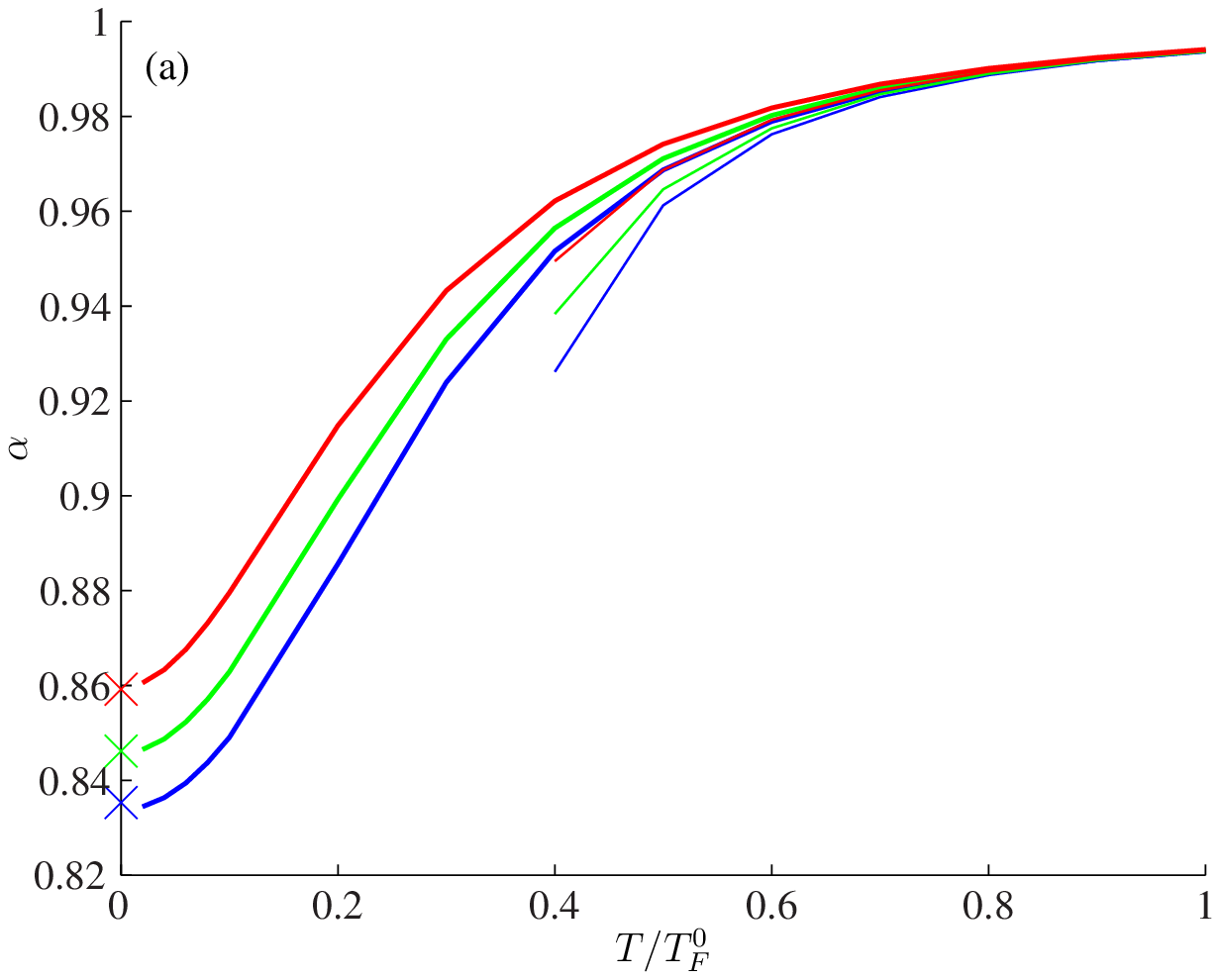}
    \includegraphics[width=3.4in]{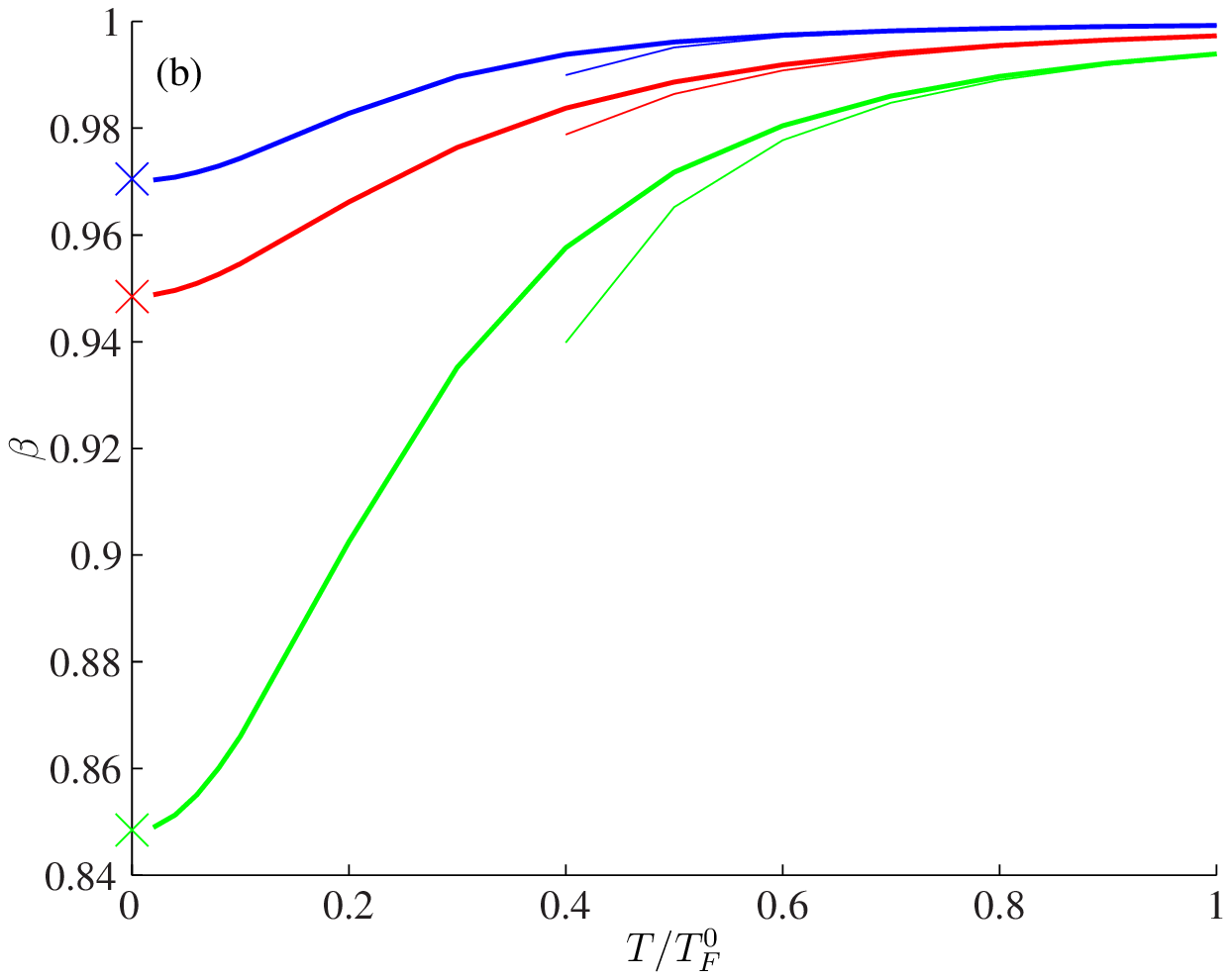} 
\caption{(Color online) Deformation of (a) momentum and (b) position density distributions.  Calculation parameters: $\lambda=0.1$ (blue/dark grey lines), $1$ (green/light grey lines) and 10 (red/grey lines). The dimensionless interaction strength is  $D_t=1$. Thin curves for $T>0.4T_F^0$ use the variational approach of \cite{Endo2010a} and crosses at $T=0$ use the variational approach of \cite{Miyakawa2008a}.
 \label{FigDeformation}}
\end{center}
\end{figure}

In Fig.~\ref{FigDeformation} we show our results for the deformation parameters  for the same parameter regime considered in Fig.~5 of Ref.~\cite{Zhang2010a}. We find our results are visually identical to their results and a quantitative comparison of data reveals a maximum relative difference of $0.11\%$. 
%\footnote{We note that we have been unable to obtain agreement with the Hartree-Fock results  in Fig.~3 of Ref.~\cite{Lin2010a}.}. 
Our results show that thermal fluctuations almost completely wash out the interaction induced deformations as the Fermi temperature is approached. We also find good agreement with the variation predictions for deformation by Miyakawa \textit{et al.} \cite{Miyakawa2008a} at $T=0$ and by Endo \textit{et al.} \cite{Endo2010a} near $T=T_F^0$ (shown in Fig.~\ref{FigDeformation}). As the finite temperature variational result is based on a Boltzmann description it is only valid in the high temperature regime $T\gg T_F^0$ and provides a useful check of the  asymptotic behavior. We note that, at the variational minimum in \cite{Endo2010a}, the kinetic energy is always equal to the high temperature limit, $3N\kt/2$, which may affect the accuracy of predictions for $\alpha$.

\subsubsection{Direct and exchange energy}\label{SecDirectExchange}
For many of the results presented in this paper we observe an appreciable difference in the behavior of thermodynamic parameters of the trapped system from the uniform gas (see \cite{Zhang2010a}). The underlying reason is the rather different role of interactions in the two systems:  For the uniform system the direct interaction term vanishes because the spatial density is uniform. In contrast, in the trapped system both direct and exchange terms contribute. To demonstrate their importance  it is useful to define their individual contributions to the system energy as [see Eq.~(\ref{Efunc})]
\begin{align}
E_D &= \frac{1}{2}\int \frac{d\x d\k}{(2\pi)^3} \Phi_D(\x)W(\x,\k),\\
E_E &= -\frac{1}{2}\int \frac{d\x d\k}{(2\pi)^3} \Phi_E(\x,\k)W(\x,\k),
\end{align}
which we shall refer to as the direct and exchange energies, respectively.

\begin{figure}[!tbh]
\begin{center} 
\includegraphics[width=3.4in]{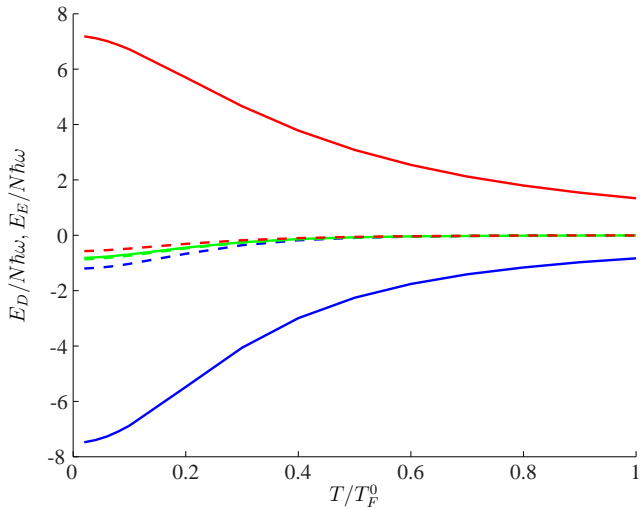} 
\caption{(Color online) Direct and exchange energy versus temperature for a dipolar Fermi gas.\label{FigDirectExchange}  Direct energy $E_D$ (solid) and exchange energy $E_E$ (dashed). Aspect ratios $\lambda=0.1$ (blue/dark grey lines), $1$ (green/light grey lines) and 10 (red/grey lines). Interaction parameter  $D_t = 1$. Note that the $E_D$ and $E_E$ results for the $\lambda=1$ case are almost indistinguishable. }
\end{center}
\end{figure}

 \begin{figure}[!tbh]
\begin{center} 
\includegraphics[width=3.4in]{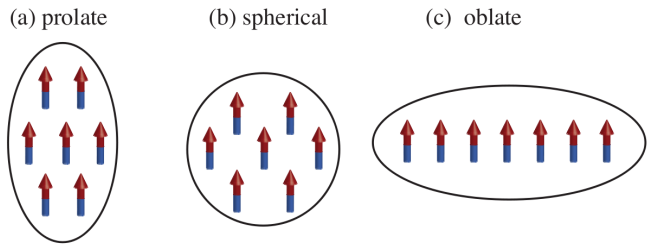} 
\caption{\label{dipgeom}(Color online) Schematic showing how the distribution of dipoles in (a) prolate, (b) spherical, and (c) oblate geometries. }
\end{center}
\end{figure}

We present results for these energies in Fig~\ref{FigDirectExchange}. We observe that the direct interaction energy is strongly effected by the trap geometry, is significantly increased in magnitude in highly anisotropic traps and can be both positive and negative. The exchange interaction energy is only slightly affected by the trap geometry and is always negative. Except for nearly spherical traps, the magnitude of $E_E$ tends to be much smaller than $E_D$.

%
%These results are easily understood as the direct and exchange interactions strongly couple to anisotropy in the  position and momentum distribution of the system, respectively.

To explain these observations we begin by considering the direct energy. The sign and strength of this quantity is controlled by the trap geometry, through its influence on the system spatial density profile: A prolate spatial density causes the attractive component of $U_{dd}$ (i.e.~interactions with dipoles in a head-to-tail configuration, see Fig.~\ref{dipgeom}(a)) to dominate and  $E_D$ is negative, and increasing in magnitude as the system becomes more prolate. An oblate spatial density causes the repulsive component of $U_{dd}$ (i.e.~interactions with dipoles in a side-by-side configuration, see Fig.~\ref{dipgeom}(c)) to dominate and $E_D$ is positive, and increasing  as the system becomes more oblate. Note that the spatial density profile has a geometry that is usually close to the trap aspect ratio $\lambda$, however   interactions cause additional distortion (as has already characterized by the parameter $\beta$). 
For example, in the spherical trap interactions will cause the spatial density to deform to be slightly prolate and $E_D$ will be slightly negative, as we see in Fig~\ref{FigDirectExchange}.

A rather similar geometric argument can be applied to the exchange interaction, but now for the momentum space distribution: 
%This is because the interaction potential $-\tilde{U}_{dd}(\k)$ governs the exchange interaction potential, which similar to the position space behavior favors a reduction in energy by
It is energetically favorable for the momentum distribution to compress along $k_\rho$ and expand along $k_z$. Interestingly for bosons, where the negative sign does not accompany the exchange term, the opposite momentum behavior would be expected.
However, any anisotropy in the momentum density arises  solely from the dipole interaction itself in all cases we consider\footnote{The isotropy of mass ensures the non-interacting momentum distribution is isotropic, however this could be changed, e.g.~by using an optical lattice to modify the effective mass in different directions. It may be possible to extend our meanfield analysis to this case, e.g.~see \cite{Baillie2009b}.} and $|E_E|$ tends to be much smaller than $|E_D|$, except in nearly spherical traps.

%
% However, an anisotropic harmonic trap will force the density to be anisotropic leading to a strong enhancement of the direct interaction. These properties are readily observed in Fig~\ref{FigDirectExchange}. 
% 
% For a spherical trap the direct and exchange energies are similar and both slightly negative, since the interactions distort the respective density distributions to be slightly prolate. In contrast, for the oblate trap ($\lambda=10$)   $E_D$ is strongly enhanced and positive, and for the prolate trap ($\lambda=0.1$) $E_D$ is also strongly enhanced, but negative.   The sign of the energy depends on geometry: A prolate spatial density causes the attractive component of $U_{dd}$ (i.e.~interactions with dipoles in a head-to-tail configuration) to dominate and  $E_D$ is negative. An oblate spatial density causes the repulsive component of $U_{dd}$ (i.e.~interactions with dipoles in a side-by-side configuration) to dominate and $E_D$ is positive. The exchange interaction energy is affected by the trap geometry, however it tends to be an order of magnitude smaller than the direct term, and is always negative.

\subsubsection{Chemical potential}
\begin{figure}[!tbh]
\begin{center}
%load cutoff;plotMu(results)
 \includegraphics[width=3.4in]{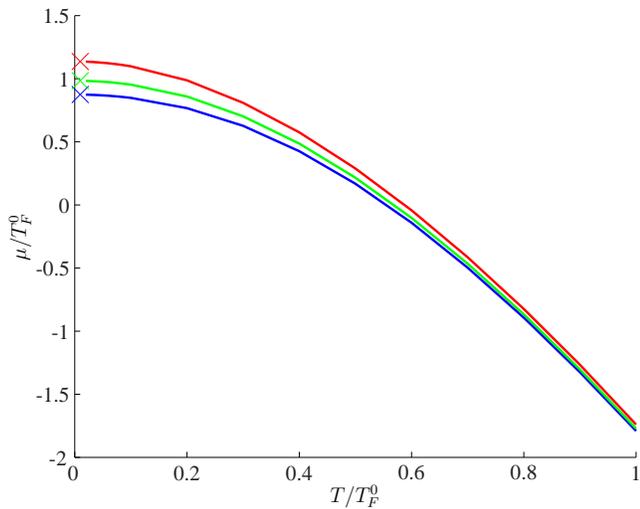}
    \caption{(Color online) Chemical potential ($\mu/k_BT_F^0$) as a function of temperature.  Calculation parameters: $\lambda=0.1$ (blue/dark grey lines), $1$ (green/light grey lines) and 10 (red/grey lines). The dimensionless interaction strength is  $D_t=1$.  Values at $T=0.01T_F^0$ shown with crosses.  \label{FigMu}}
\end{center}
\end{figure}

We present results for the low temperature chemical potential of the trapped dipolar gas in Table \ref{chempottable} for the main system parameters considered in this paper. We consider the ratio  $\mu/k_BT_F^0$, which gives us a quantitative measure of the effect of interactions on the Fermi temperature of the interacting system.
 In practice we evaluate these results at the small but finite temperature of $T=0.01T_F^0$.  We do this because the lowest temperature we can solve for is limited by the ability of the computational grids we use to resolve the sharp Fermi surface. However, as shown in Fig.~\ref{FigMu}, the chemical rapidly saturates as $T\to0$ and the values calculated at $T=0.01T_F^0$ should be very close to the $T=0$ value.

\begin{table}[h]
\begin{center} {
\begin{tabular}{l|ccc}
%
%\multicolumn{4}{c}{Chemical potential: $\mu/k_BT_F^0$}\\
\hline\hline
& \multicolumn{3}{c}{$\lambda$}\\
$D_t$  \hfill & 0.1  &     1       &         10 \\
 \hline \hline
0.5    &      0.95  &  1.00  &  1.07 \\
  1      &      0.87  &  0.98  &  1.14\\
  2      &       N/A &      N/A &          1.24\\
\end{tabular} }
\end{center}
\caption{\label{chempottable} Low temperature chemical potential $\mu/k_BT_F^0$ for the system parameters considered in this paper. Chemical potential is evaluated from our Hartree-Fock solution at a temperature of $T=0.01T_F^0$. }
\end{table}

The behavior of the chemical potential has been considered for the uniform gas in \cite{Zhang2010a}. In that work it was found that increasing the dipolar interaction strength suppressed $\mu$. In the trapped system we observe that prolate trapping geometry can enhance this suppression, while an oblate geometry can instead cause the chemical potential to increase with increasing interaction strength.

\subsubsection{Entropy: Mechanism for mechanical cooling}
It is of interest to know how the entropy of the system depends on the other parameters, such as the trapping potential and interaction strength. We can directly calculate the entropy from the Wigner function as
\begin{align}
S=&-\int \frac{d\x d\k}{(2\pi)^{3}} \Big\{W(\x,\k)\ln W(\x,\k)\nonumber \\
&+[1-W(\x,\k)]\ln[1-W(\x,\k)]\Big\}.\label{EqEntropy}
\end{align}
\begin{figure}[!tbh]
\begin{center}
%load cutoff
%plotTherm(results, 'entropy', 'S/N\kb', false, @(t)4+log(6)+3*log(t))
%plotThermvsNI(results, 'entropy', 'S')
%inkscape
    \includegraphics[width=3.4in]{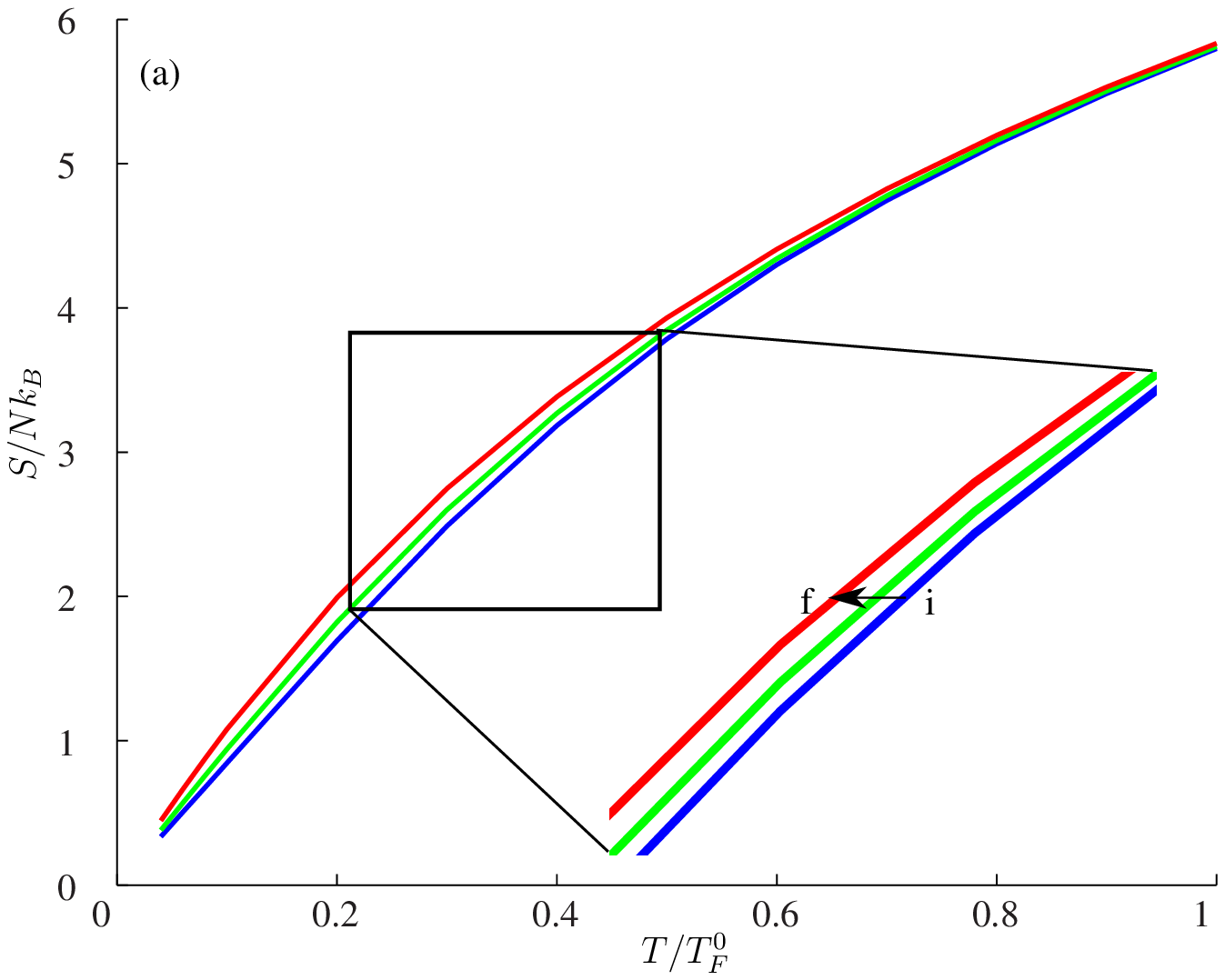}  
\includegraphics[width=3.4in]{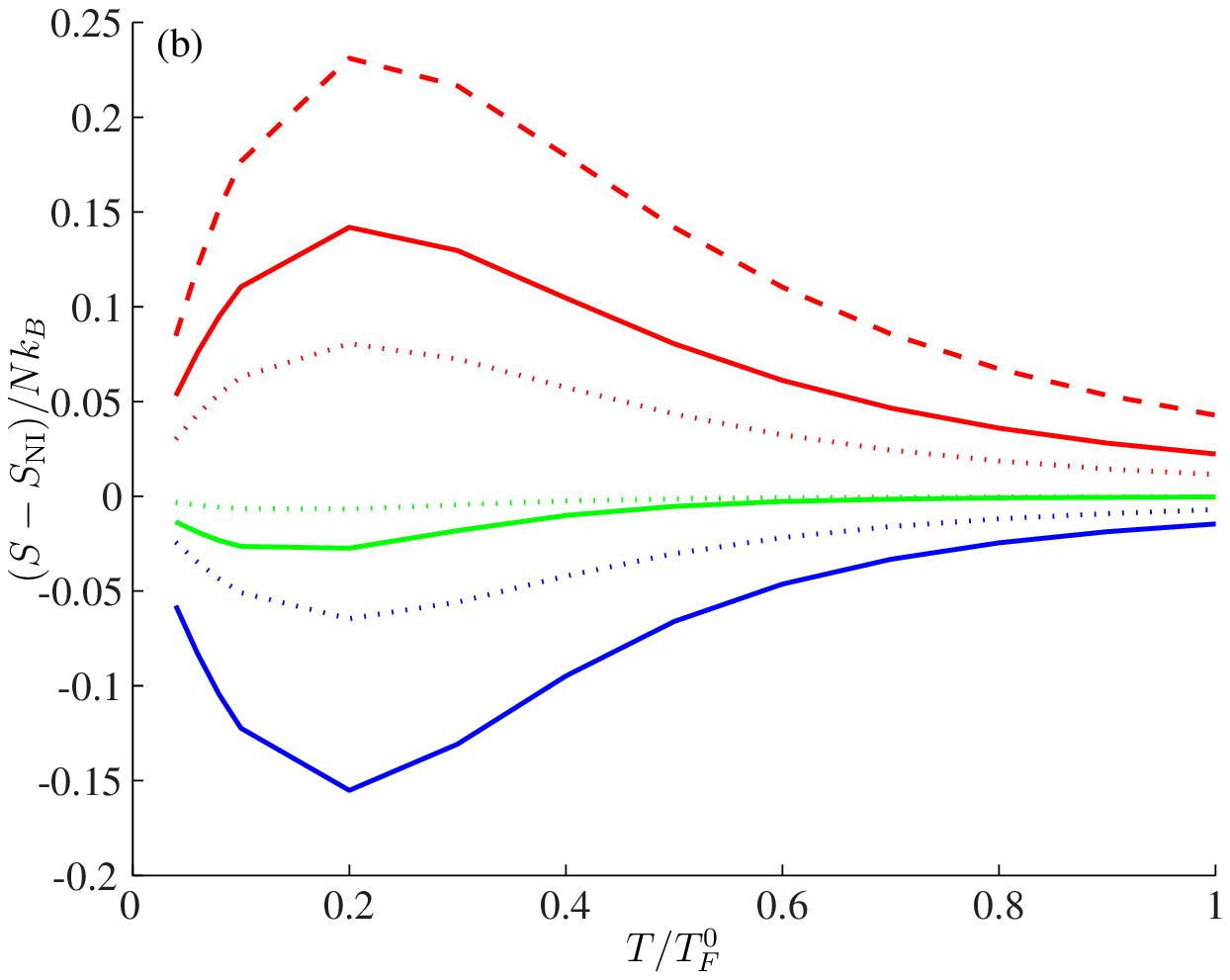} 
\caption{(Color online) Entropy versus temperature for a dipolar Fermi gas. (a) Entropy versus temperature. (b) Difference in entropy from the ideal Fermi gas.\label{FigEntropy}  Inset shows a magnification of the entropy curves with a adiabatic process   from initial condition (i) to final condition (f) indicated.
Aspect ratios  $\lambda=0.1$ (blue/dark grey lines), $1$ (green/light grey lines) and 10 (red/grey lines); Interaction parameter $D_t = 0.5$ (dotted), $D_t = 1$
(solid), $D_t = 2$ (dashed).}
\end{center}
\end{figure}
In Fig.~\ref{FigEntropy}(a) we show results for the entropy versus temperature for three different trap geometries with constant interaction strength. In Ref.~\cite{Zhang2010a} the entropic behavior of the uniform gas was studied and in their numerical simulations they found that as the strength of the dipolar interactions were increased the value of entropy decreased (at fixed temperature). We  find even richer behavior in the trapped system.
 When the position space density is prolate (i.e.~$\lambda<1$) interactions tend to decrease the entropy  (most easily seen by considering the difference in entropy form the non-interacting case shown in Fig.~\ref{FigEntropy}(b)). In contrast, for an oblate density distribution the behavior is the opposite of the uniform case and the entropy  increases with increasing interaction strength. 
 
We have also computed the cases shown in Fig.~\ref{FigEntropy}(b) but neglecting exchange interactions (see Sec.~\ref{SecHthry}), and find that universally this increases the value of entropy by a small amount (relative to results including exchange). This demonstrates  that the entropic shifts we observe  are dominated by direct interactions, consistent with the energetic observations made in Sec.~\ref{SecDirectExchange}.

Interestingly, for all cases shown in Fig.~\ref{FigEntropy}(b) the difference arising from interactions   is maximum at a temperature of  $T\approx0.2T_F^0$, with little dependence on $D_t$ or $\lambda$. This coincides with the temperature at which we observe the most rapid change in the deformation parameters (see Fig.~\ref{FigDeformation}).   
Thus, $0.2T_F^0$ appears to set the appropriate temperature scale for experiments to achieve in order to observe the onset of strong dipolar effects in thermodynamic properties. Currently this is approximately an order of magnitude colder than the coldest reported temperature for polar molecules.

We observe the interesting feature that as the trap ratio $\lambda$ increases the entropy curves shift to lower temperatures.  Thus, if the trap geometry is changed adiabatically (constant entropy) then the system temperature should decrease (see process indicated in inset to Fig.~\ref{FigEntropy}(a)). 
This realizes a mechanical cooling process for the system induced by squeezing the dipolar gas along the polarization direction.  Overall this scheme offers rather limited scope for cooling, but as there is  considerable work  underway in experiments to reduce the temperature of polar molecules to the degenerate regime, this scheme might be a useful final stage in the cooling sequence.

\subsubsection{Heat capacity}
\begin{figure}[!tbh]
\begin{center}
%plotTherm(results, 'heatcapacity', 'C/N\kb', false,@(t)3)
%plotThermvsNI(results, 'heatcapacity', 'C')
\includegraphics[width=3.4in]{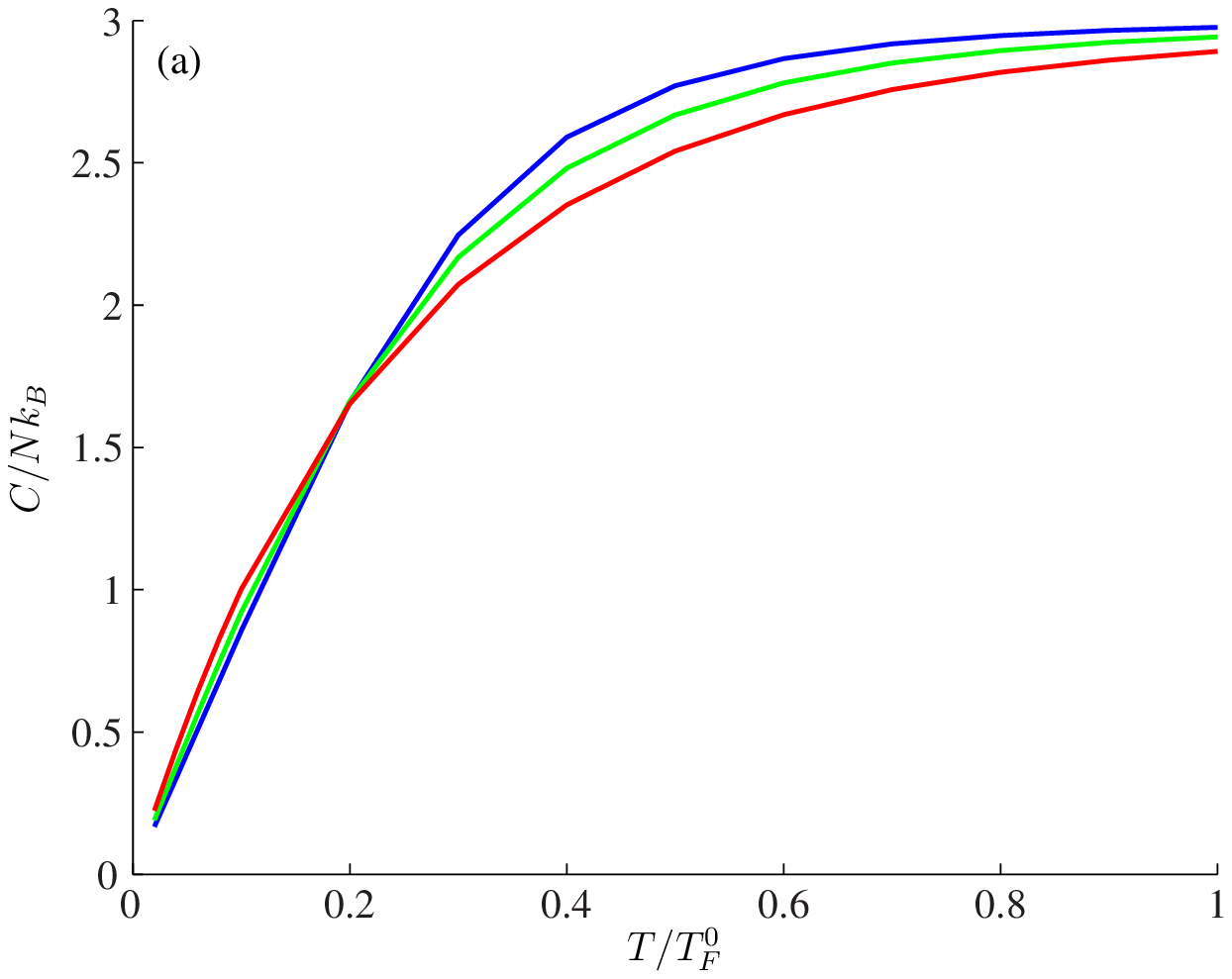}
\includegraphics[width=3.4in]{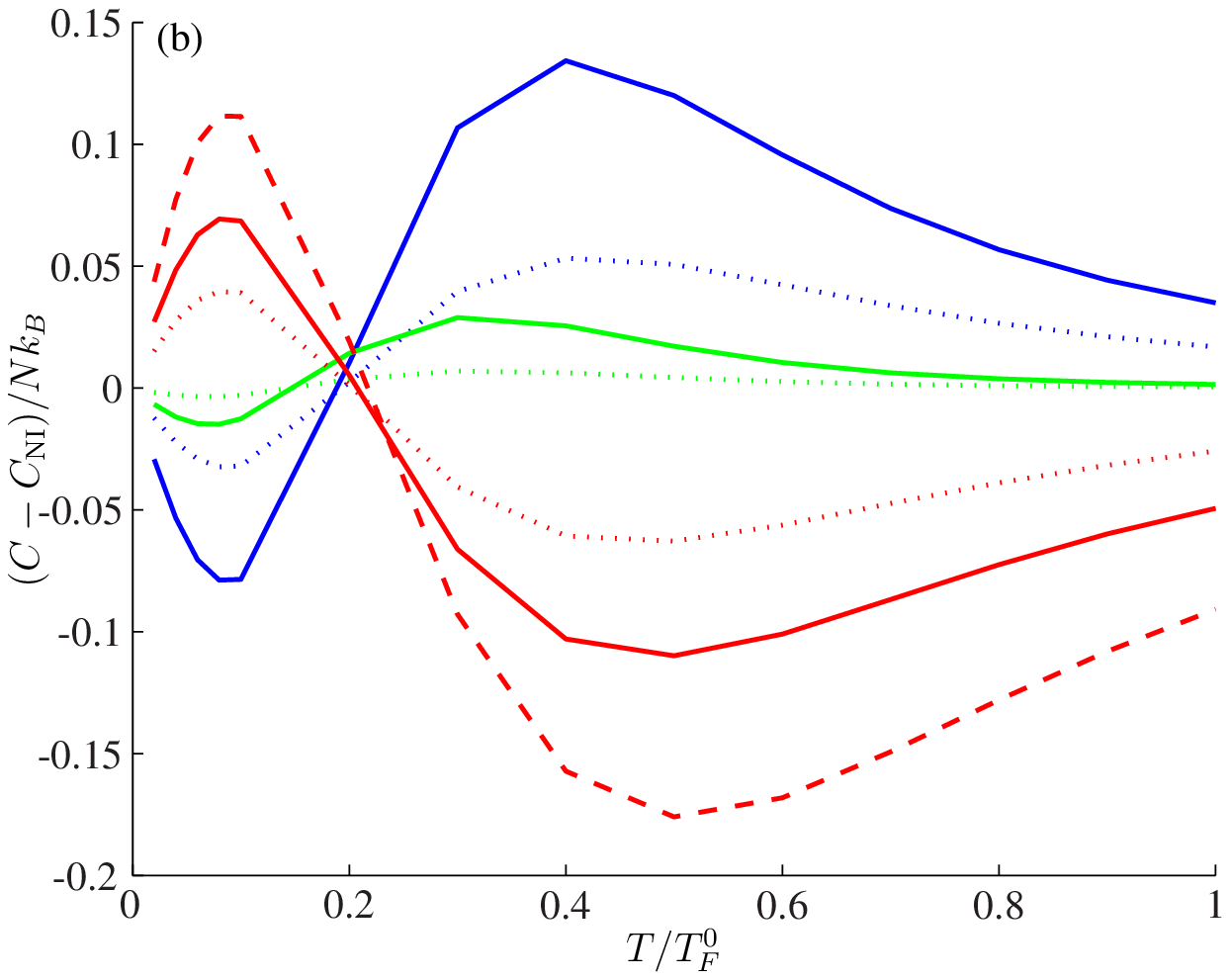} 
\caption{(Color online) Heat capacity versus temperature for a dipolar Fermi gas.\label{FigHeatCapacity} (a) Heat capacity versus temperature. (b) Change in the interacting heat capacity from the ideal gas heat capacity. Aspect ratios $\lambda=0.1$ (blue/dark grey lines), $1$ (green/light grey lines) and 10 (red/grey lines). Interaction parameter $D_t = 0.5$ (dotted), $D_t = 1$
(solid), $D_t = 2$ (dashed).}
\end{center}
\end{figure}

The heat capacity can be evaluated as
\begin{equation}
C=\left(\frac{\partial E}{\partial T}\right)_N = T\left(\frac{\partial S}{\partial T}\right)_N,\label{heatcapacity}
\end{equation}
i.e.~either through the energy functional (\ref{Efunc}) or the entropy (\ref{EqEntropy}). We have applied both methods and find they agree. These definitions require us to take a numerical derivative, which we do by finite difference using two solutions that differ in temperature by $\Delta T = 10^{-4} T$. Our results for the heat capacity are shown in Fig.~\ref{FigHeatCapacity}.

The  behavior seen in the heat capacity is easily understood from our entropy results [see Fig.~\ref{FigEntropy}(b), and Eq.~(\ref{heatcapacity})]. In particular the crossover of the curves in Fig.~\ref{FigHeatCapacity}(a) (i.e.~where the difference curves cross the horizontal axis in Fig.~\ref{FigHeatCapacity}(b)) occurs at $T\approx0.2T_F^0$ where the peak in entropy difference was observed in Fig.~\ref{FigEntropy}(b).

\subsection{Dipolar gas coherence}
Developments in experimental techniques (e.g.~see \cite{Greiner2005a,Schellekens2005a,Folling2005a,Rom2006a,Donner2007a}) have enabled the measurement of correlations in ultra-cold gases. Such measurements provide useful many-body information (e.g.~see \cite{Altman2004a,Toth2008a}).

The first order coherence of the gas is described by the two point correlation function 
\begin{equation}
G^{(1)}(\x,\xd)=\langle \hat{\psi}^\dagger(\x)\hat{\psi}(\xd)\rangle,
\end{equation}
(e.g.~see \cite{Naraschewski1999a}) where $\hat{\psi}(\x)$ is the fermionic quantum field operator. 
Introducing center of mass [$\R=\frac{1}{2}(\x+\xd)$] and relative [$\r=\x-\xd$] coordinates,  the correlation function relates directly to the Wigner function as
\begin{equation}
G^{(1)}(\R,\r)=\int \frac{d\mathbf{k}}{(2\pi)^3}\,W(\R,\k)e^{i\mathbf{k}\cdot\r}.\label{G1Wigner}
\end{equation}
While it is possible to experimentally measure the first order correlation function at two specific points in a trapped gas \cite{Donner2007a}, it is useful to consider a volume-averaged correlation function that eliminates the dependence on the center of mass coordinate (e.g.~see \cite{Greiner2005a,Schellekens2005a,Folling2005a,Rom2006a}). Motivated by this we define the  volume-averaged first order correlation function
\begin{equation}
g^{(1)}(\r)=\frac{\int d\R \,G^{(1)}(\R,\r)}{\int d\R\,\sqrt{ n\!\left(\R+\frac{1}{2}\r\right)n\!\left(\R-\frac{1}{2}\r\right)}},
\end{equation}
which we have normalized by the density (see Eq.~(2.20) of \cite{Naraschewski1999a}) so that $g^{(1)}(\mathbf{0})=1$.

Using Hartree-Fock factorization we can also consider higher order correlation functions, such as  the second order expression  
\begin{align}
G^{(2)}(\x,\xd)&=\langle\hat\psi^\dagger(\x)\hat\psi^\dagger(\xd)\hat\psi(\xd)\hat\psi(\x)\rangle,\\
&=n(\x)n(\xd)- |G^{(1)}(\x,\xd)|^2, 
\end{align}
which relates to the system density correlations.
The normalized and volume-averaged second order correlation function is given by
\begin{equation}
g^{(2)}(\r)=1-\frac{\int d\R\,|G^{(1)}(\R,\r)|^2}{\int d\R\,\,n\!\left(\R+\frac{1}{2}\r\right)n\!\left(\R-\frac{1}{2}\r\right)}.
\end{equation}

\begin{figure}[!tbh]
\begin{center}
%plotCoherence
    \includegraphics[width=3.4in]{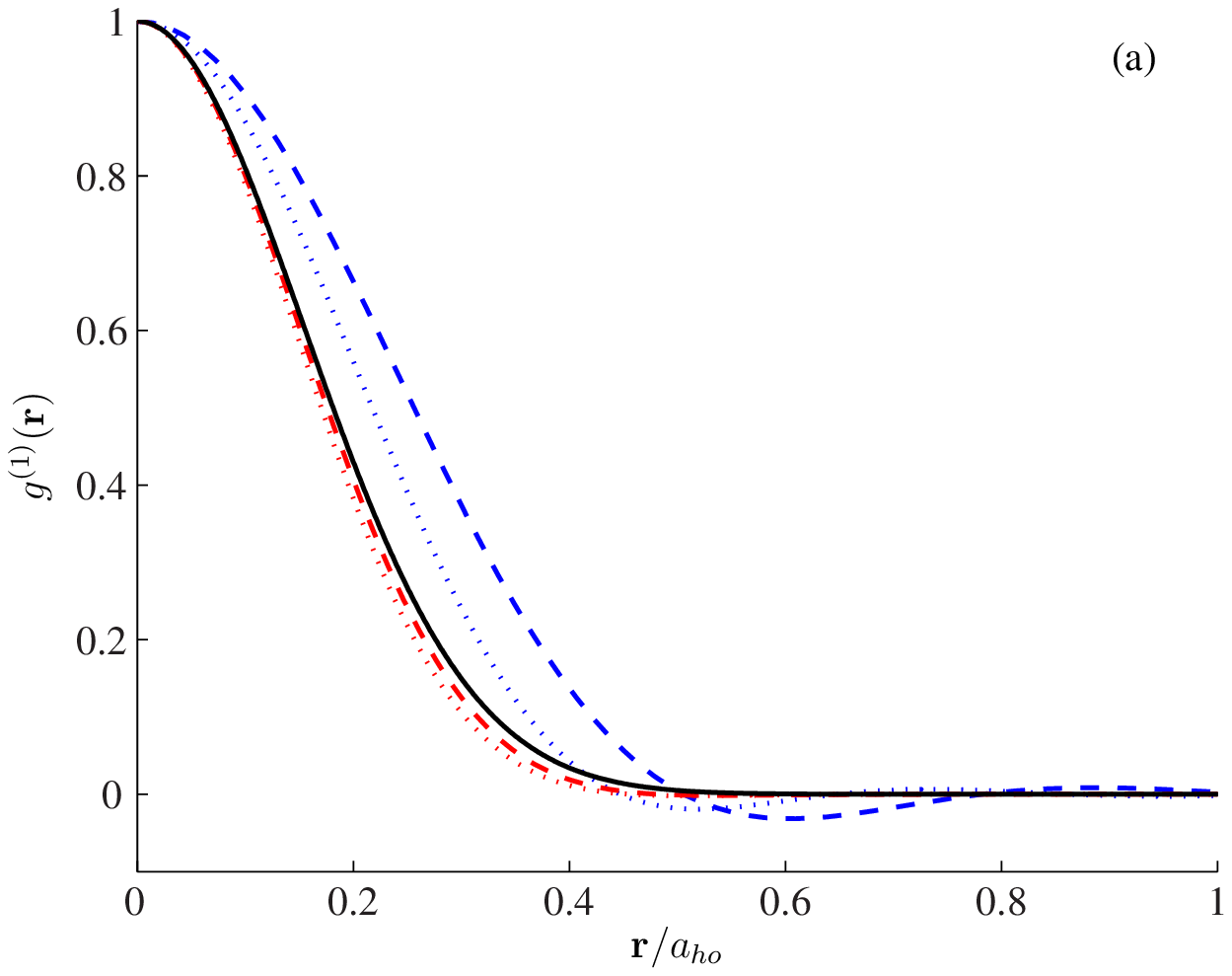}
\includegraphics[width=3.4in]{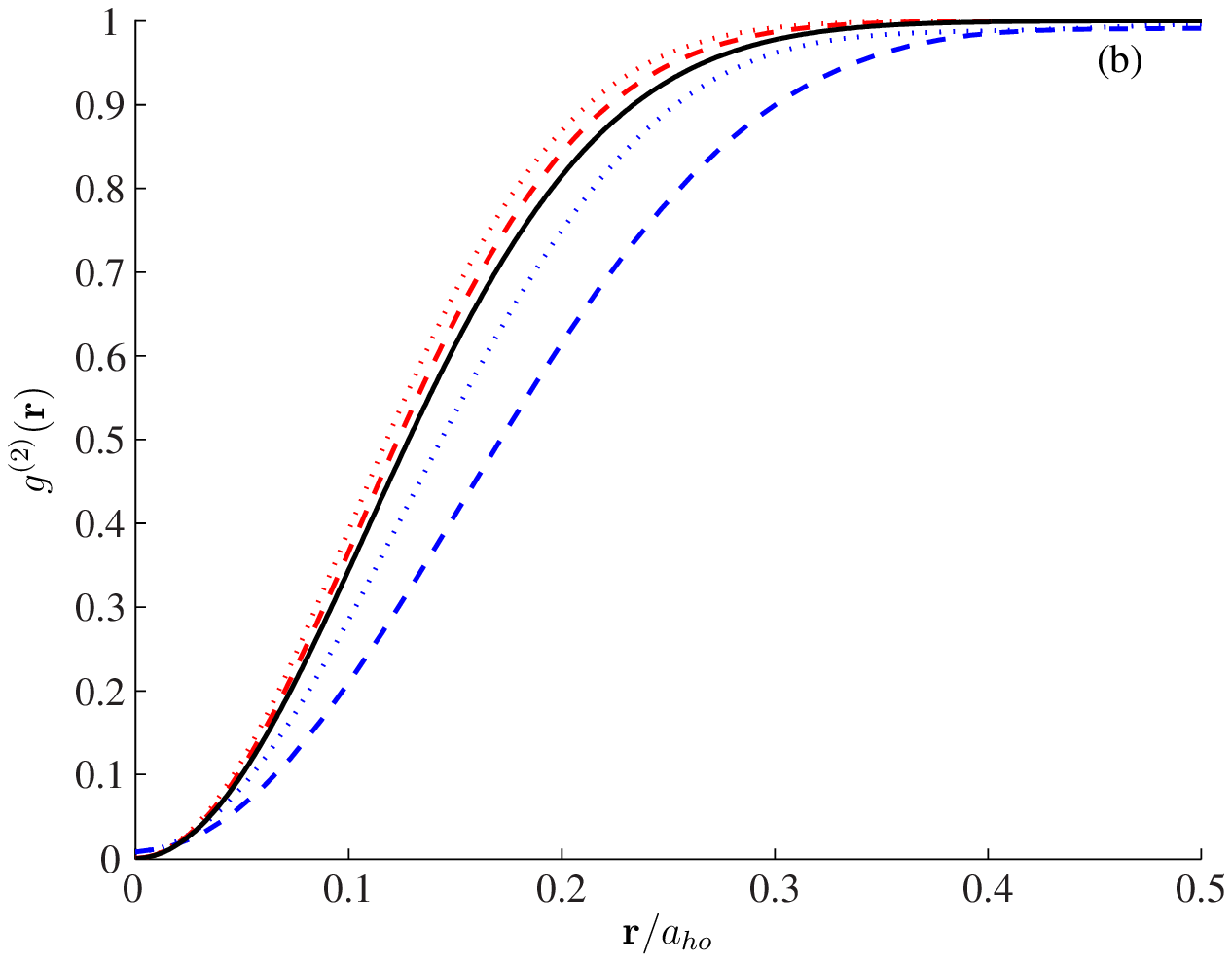}  
\caption{(Color online) Normalized volume averaged correlation functions for the dipolar Fermi gas. (a) The first order correlation function $g^{(1)}(\r)$ and high temperature limit (black curve) $e^{-\pi |\br|^2/\lambda_{dB}^2}$. (b) The second order correlation function  $g^{(2)}(\r)$ and high temperature limit $1-e^{-2\pi |\br|^2/\lambda_{dB}^2}$. Results given along the $x$ axis (dashed) and along the $z$ axis (dotted) for aspect ratio $\lambda=1$ (other aspect ratios show similar behaviour) and $D_t=1$.  Results for $T=0.01T_F^0$ (blue/grey) and $T=0.5T_F^0$ (red/light grey). $T=0.5T_F^0$ is used for the high temperature limits. Results for both aspect ratios are indistinguishable from their  high temperature limit at $T=T_F^0$.
\label{Fig_g1g2} }
\end{center}
\end{figure}

Results for the $g^{(1)}(\r)$ and  $g^{(2)}(\r)$ are shown in Fig.~\ref{Fig_g1g2}. The low temperature case of $g^{(1)}(\r)$ considered in Fig.~\ref{Fig_g1g2}(a) exhibits asymmetry: the length scale for the decay of coherence is shorter along the $z$ direction than the $x$ (radial) direction. For higher temperatures, this feature is washed out by thermal fluctuations and the coherence becomes isotropic. We have verified that the low temperature asymmetry in the coherence arises from exchange interactions. To do this we set $\Phi_E(\x,\k)$ to zero in our calculations (but still retaining the direct interactions, see Sec.~\ref{SecHthry}) and found that the correlation functions are isotropic.  
 
 The $g^{(2)}(\r)$ correlation function reveals the expected anti-bunching behavior for fermions (see  Fig.~\ref{Fig_g1g2}(b)), i.e. $g^{(2)}({\bf 0})=0$. We also observe asymmetry in these second order correlations at low temperatures similar to that seen for the first order coherence function.

%
% 
%We can now From Eq.~(\ref{G1Wigner}) we immediately see that 
%\begin{equation}
%\int d\mathbf{R}\,G^{(1)}(\mathbf{R},\mathbf{r})=\int d\k\,n(\k)e^{i\k\cdot\r},\label{G1widthMtmwidth}
%\end{equation}
% so the width  of the coherence function is directly related to the momentum distribution.
% 

\begin{figure}[!tbh]
\begin{center}
%plotCoherenceLength
\includegraphics[width=3.4in]{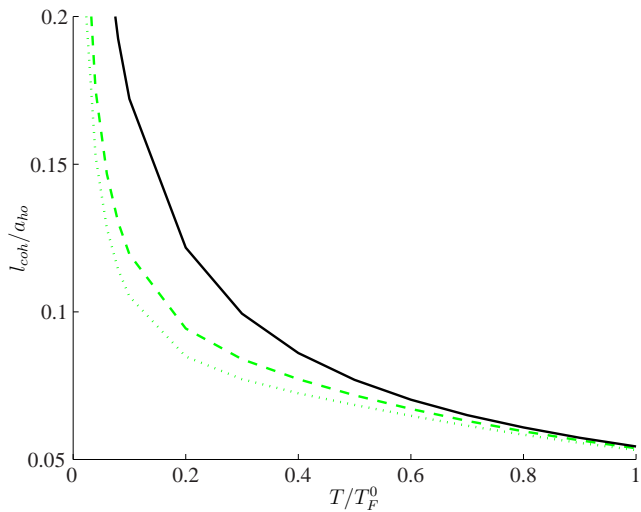}
\caption{(Color online) Coherence length of a dipolar Fermi gas as a function of temperature: $l_{coh,x}$ (dashed line), $l_{coh,z}$ (dotted line). 
Results are given for $D_t=1$ and aspect ratio $\lambda=1$. Aspect ratios  $\lambda=0.1$ and  $\lambda=10$ are similar. The black curve shows  $\lambda_{dB}/\sqrt{8\pi}$ for reference.
\label{FigCoherence}
}
\end{center}
\end{figure}

 To characterize the first order coherence more thoroughly we follow \cite{Barnett2000a} and define the coherence length in the  $x$-direction ($l_{coh,x}$) as
\begin{equation}
\left(l_{coh,x}\right)^{2}=\frac{\int d\mathbf{r}\int d\mathbf{R}|G^{(1)}(\mathbf{R},\mathbf{r})|^{2}x^{2}}{2\int d\mathbf{r}\int d\mathbf{R}|G^{(1)}(\mathbf{R},\mathbf{r})|^{2}},\label{EqBarnettCoherence}
\end{equation}
and similarly for the other directions (Note: $x$ is a component of the relative coordinate). At high temperatures we have the limiting (Boltzmann) behavior $l_{coh}\to\lambda_{dB}/\sqrt{8\pi}$ (independent of direction), with $\lambda_{dB}=h/\sqrt{2\pi mk_BT}$.  The results for the coherence length are shown in Fig.~\ref{FigCoherence}, and reveals the dependence of the coherence anisotropy on temperature.

% As $T\to0$ our results indicate that the correlation length appears to diverge.  To understand this we first note that from Eq.~(\ref{G1Wigner}) we have
% \begin{equation}
%\int d\mathbf{R}\,G^{(1)}(\mathbf{R},\mathbf{r})=\int d\k\,n(\k)e^{i\k\cdot\r},
%\end{equation}
%and so that the variance in the Fourier transformed momentum distribution corresponds to the numerator of our expression for the coherence length. In the zero temperature limit the momentum distribution becomes a step function, for which the Fourier transform has a divergent variance. However, beyond the semiclassical approximation (which breaks down when $k_BT\sim\hbar\omega$) the momentum distribution is rounded off by the trapping potential  and will  approach a finite value. 

\subsection{Hartree results}\label{SecHthry}
The results presented so far are computationally demanding and it is desirable to produce a simpler theory. 
We have observed that, except in the nearly spherical traps, the exchange interaction is typically much smaller than the direct interaction. Thus neglecting the exchange (Fock) term would seem to be a reasonable approximation in anisotropic traps, and we can use the simplified Hartree dispersion
\begin{equation}
\epsilon(\x,\k)=\frac{\hbar^2k^2}{2m}+U(\x)+\Phi_D(\x).\label{eqn_Hepsilon}
\end{equation}
Since the $k$ dependence is of a simple isotropic form it can be integrated out to obtain
\begin{align}n(\x)&=\int \frac{d\k}{(2\pi)^3}W(\x,k),\\
&=\frac{1}{\lambda_{dB}^{3}}\zeta^-_{3/2}\left(e^{\left[\mu-V_{\rm{eff}}(\x) \right]/k_BT}\right),\label{nloc}\end{align}
 where 
 %$\lambda_{dB}=h/\sqrt{2\pi mk_BT}$, and 
 $\zeta_\nu^-(z)=\sum_{k=1}^\infty (-1)^{k-1}z^k/k^\nu$ is the Fermi function, and $V_{\rm{eff}}(\x) =U(\x)+\Phi_D(\x)$ is the effective potential. Thus the spatial density can be obtained self-consistently using Eq.~(\ref{nloc}), without needing to construct $W(\x,\k)$ or $\epsilon(\x,\k)$.
 
To obtain some information about the applicability of the Hartree theory, we compare its predictions to the full Hartree-Fock calculations for a range of thermodynamic parameters in Table \ref{Htable}. We observe that agreement between the two theories is best for highly anisotropic traps, small interaction strength, and high temperatures. Of course we note that in all cases the Hartree theory would predict $\alpha=1$ (since momentum distortion is solely from exchange interactions) and that the coherence properties of the gas would be isotropic.  
 
\begin{widetext}
\begin{center} 
\begin{table}[h]
{ 
\begin{tabular}{ll|cc|cc|cc||cc|cc|cc|}
\hline\hline
&& \multicolumn{6}{c||}{$T=0.01T_F^0$} & \multicolumn{6}{c|}{$T=0.5T_F^0$}\\[6pt]
\hline
&& \multicolumn{2}{c}{$\mu/k_B T_F^0$} & \multicolumn{2}{c}{$\beta$} & \multicolumn{2}{c||}{$E_D/N\hbar\omega$}&\multicolumn{2}{c}{$\mu/k_B T_F^0$} & \multicolumn{2}{c}{$\beta$} & \multicolumn{2}{c|}{$E_D/N\hbar\omega$}\\
$\lambda$ &$D_t$  \hfill &HF & H &HF & H & HF & H \hfill &HF & H &HF & H & HF & H\\
 \hline
0.1 & 0.5 &  0.945 & 0.947 & 0.988 & 0.989 & -3.11 & -3.09 &  0.194 & 0.194 & 0.998 & 0.998 & -1.07 & -1.07\\
 & 1 &  0.874 & 0.887 & 0.970 & 0.972 & -7.50 & -7.20 &  0.168 & 0.169 & 0.996 & 0.996 & -2.26 & -2.25\\\hline
0.25 & 0.5 &  0.954 & 0.957 & 0.966 & 0.966 & -2.54 & -2.53 &  0.198 & 0.199 & 0.994 & 0.994 & -0.88 & -0.88\\
 & 1 &  0.896 & 0.908 & 0.921 & 0.924 & -6.02 & -5.80 &  0.177 & 0.178 & 0.988 & 0.988 & -1.84 & -1.83\\\hline
0.5 & 0.5 &  0.971 & 0.973 & 0.941 & 0.941 & -1.61 & -1.60 &  0.206 & 0.206 & 0.990 & 0.990 & -0.55 & -0.55\\
 & 1 &  0.931 & 0.941 & 0.875 & 0.880 & -3.86 & -3.72 &  0.192 & 0.193 & 0.979 & 0.979 & -1.14 & -1.14\\\hline
1 & 0.5 &  0.996 & 0.998 & 0.922 & 0.923 & -0.20 & -0.19 &  0.217 & 0.218 & 0.986 & 0.986 & -0.02 & -0.02\\
 & 1 &  0.983 & 0.992 & 0.848 & 0.853 & -0.82 & -0.77 &  0.216 & 0.217 & 0.972 & 0.972 & -0.07 & -0.07\\\hline
2 & 0.5 &  1.025 & 1.027 & 0.923 & 0.924 & 1.44 & 1.44 &  0.231 & 0.232 & 0.985 & 0.985 & 0.60 & 0.60\\
 & 1 &  1.043 & 1.050 & 0.859 & 0.863 & 2.49 & 2.47 &  0.243 & 0.244 & 0.972 & 0.972 & 1.15 & 1.15\\\hline
4 & 0.5 &  1.052 & 1.053 & 0.942 & 0.942 & 2.85 & 2.84 &  0.244 & 0.244 & 0.989 & 0.989 & 1.15 & 1.15\\
 & 1 &  1.094 & 1.100 & 0.898 & 0.900 & 5.16 & 5.10 &  0.268 & 0.269 & 0.979 & 0.979 & 2.20 & 2.20\\\hline
10 & 0.5 &  1.074 & 1.076 & 0.969 & 0.970 & 4.01 & 4.00 &  0.255 & 0.255 & 0.994 & 0.994 & 1.62 & 1.62\\
 & 1 &  1.136 & 1.141 & 0.949 & 0.950 & 7.20 & 7.12 &  0.288 & 0.289 & 0.989 & 0.989 & 3.08 & 3.08\\\hline
\end{tabular} }

\caption{\label{Htable} Comparison of Hartree (H) and Hartree-Fock (HF) predictions for thermodynamic parameters.  }
\end{table}
\end{center}
\end{widetext}

\section{Conclusions and outlook}

  In this paper we have developed a finite temperature meanfield treatment of a single component  trapped  Fermi gas with dipole-dipole interactions. We have discussed the details of our numerical implementation to assist others in implementing this theory, and have compared to the numerical results of  Zhang and Yi \cite{Zhang2010a} and variational theories to validate our calculations.
    
  We have considered a range of thermodynamic properties of the system. We have found that the long-ranged and anisotropic interaction causes many system properties to depend on the trap geometry, beyond simple scaling with the (geometric) mean trap frequency. Interestingly we find that dipolar interactions  allow the system to be cooled as the harmonic confinement is used to squeeze the system towards an oblate geometry. 
  
We have constructed the first and second order correlation functions within the Hartree-Fock approximation, and have evaluated these in the volume-averaged form relevant to experiments. We find that the exchange interactions give rise to anisotropy in these correlation functions, clearly revealed in the coherence length of the system. 

Importantly, our results show the sensitivity of many of the features of the dipolar Fermi gas to temperature.  In particular, over a wide range of parameters and observables we find that the strongest effects of the dipolar interactions are apparent for $T\lesssim 0.2T_F^0$. Additionally, we find that many of the qualitative predictions made for the uniform gas are strongly modified in the trapped system by strong direct interaction effects, particularly in oblate traps. Finally, we have investigated a  simpler Hartree theory that  avoids the technical complexities associated with evaluating the exchange interactions, and is much faster to calculate. We have validated its applicability and accuracy for studying thermodynamic properties in anisotropic trapping potentials where the exchange effects are less significant.

With current fermionic polar molecule experiments progressing towards degeneracy, and suggestions that very flat traps will be necessary to reduce inelastic collision processes \cite{Ni2010a}, these results will be useful to understand the system behavior and to perform accurate thermometry.   
 
 \section*{Acknowledgements}
 We would like to acknowledge discussions with J.N.~Zhang, S.~Yi, C.H.~Lin, D.W.~Wang, and P.S.~Julienne, and thank J.N.~Zhang for providing   data to compare against. This work is supported by Marsden contract 09-UOO-093 and the New Zealand Tertiary Education Commission TADS. We also acknowledge support from the New Zealand Foundation for Research, Science and Technology contract NERF-UOOX0703.

\appendix
\section{Numerical methods\label{s:numericalmethods}}

\subsection{Choice and use of  computational grids: underlying quadratures}
 We represent the Wigner function and the Hartree-Fock dispersion as four dimensional arrays given by
 \begin{align}
     W_{qrst}&\equiv W(\rho_q,z_r,k_{\rho_s},k_{z_t}),\\
 \epsilon_{qrst}&\equiv \epsilon(\rho_q,z_r,k_{\rho_s},k_{z_t}).
 \end{align}
 We have chosen the numerical grids $\{\rho_q,z_r,k_{\rho_s},k_{z_t}\}$ to provide an efficient representation of these functions making use of their known symmetry. 
 Additionally, the grid allows us to implement numerical quadrature, which in the standard form is
  \begin{equation}
 \int_a^b dx\,w(x)f(x) \approx\sum_{j=1}^Nw_jf(x_j),\label{quad}
 \end{equation}
 where the roots   ($\{x_j\}$),  weight function ($w(x)$) and weights ($w_j$) define the quadrature. 
In the following subsections we introduce the two types of numerical grids we use in detail and discuss their related quadrature.

\subsubsection{Axial grids}
 In the axial direction we use the uniformly spaced position grid (i.e.~the  well-known grid for the discrete Fourier transform)
 \begin{align}
 z_j&=\left(j-\frac{1}{2}\right)\Delta_z,\quad &j=1,\ldots,N_{z},
 \end{align}
 where $\Delta_z$ is the point-spacing.
We  restrict this grid to positive values because of the assumed even symmetry of the functions about $z=0$. 
 We take the  grid  for the Fourier transformed variable  to be
 \begin{align}
 F_{z,j}&=\frac{\pi}{N_z\Delta_z} \left(j-\frac{1}{2}\right) ,\quad &j=1,\ldots,N_{z}.
 \end{align}

Using the trapezium rule with the discrete Fourier grid has been shown to be a Gauss-Chebychev quadrature of the first kind \cite{Muckerman1990a}, i.e.~ quadrature with weight function $w(z)=1$ and weights $w_i^{(z)}=\Delta_z$. A similar quadrature exists in  $F_z$ space with $w_i^{(F_z)}= {\pi}/{N_z\Delta_z}$.
 
We now consider how to perform the Fourier transformation.  As   the function to be transformed  is even we need only implement a cosine transform, i.e.~
 \begin{align}
 \tilde{f}(F_z)&=\int_{-\infty}^{\infty}dz\,{\cos(F_zz)}{}f(z),\\ 
\Rightarrow  \tilde{f}(F_{z,i})&\approx \sum_jT_{ij}f(z_j),\label{quadDCT}
 \end{align}
 where we have introduced the transformation matrix
\begin{equation}
T_{ij}  =2\Delta_z\cos\left[\frac{\pi}{N_z}\left(i-\frac{1}{2}\right)\left(j-\frac{1}{2}\right)\right].\label{TDCTfwd}
\end{equation} 
To obtain (\ref{quadDCT}) we have used the quadrature rule, and assumed even symmetry of the function $f(z)$. Similarly the inverse transform from   $\tilde{f}(F_{z,j})\to f(z_i)$ is given by the matrix
\begin{equation}
(T^{-1})_{ij}  =\frac{1}{ N_z\Delta_z}\cos\left[\frac{\pi}{N_z}\left(i-\frac{1}{2}\right)\left(j-\frac{1}{2}\right)\right].
\end{equation}  
Finally, we note that while we have cast this argument in terms of performing quadrature, we of course have obtained the standard discrete cosine transformation. Thus, to within  constant prefactors, the forward and inverse transform are the so called DCT-IV variant of the discrete cosine transformation available in some standard  FFT libraries (e.g.~see \cite{Frigo2005a}). 

For the case of the axial momentum variable ${k_z}$, the grid is similarly given by
 \begin{align}
k_{z_j}&=\left(j-\frac{1}{2}\right)\Delta_{k_z} ,\quad &j=1,\ldots,N_{{k_z}},
 \end{align}
 where $\Delta_{k_z}$ is the spacing between momentum points.
  The above discussion regarding the quadrature of quantities on the  position grid immediately applies to the    
 momentum case with the replacements $N_z\to N_{{k_z}}$ and $\Delta_z\to \Delta_{{k_z}}$.

\subsubsection{Radial grids}
 In the radial direction it is convenient to introduce grids based on the roots of the Bessel function to allow an efficient Hankel transform (i.e.~radial Fourier transformation) to be performed (also see  \cite{Ronen2006a}).  For the position variable we use a numerical grid of $N_\rho$ points covering the region $(0,R)$ given by
 \begin{align}
 \rho_j&=\frac{\alpha_jR}{\alpha_{N_{\rho}+1}},\quad &j=1,\ldots,N_{\rho},
 \end{align}
 where $\alpha_j$ is the $j$th root of the Bessel function $J_0(z)$.  
 We take the respective grid for the  Fourier transformed variables to be\begin{align}
 F_{\rho,j}&=\frac{\alpha_j}{R},\quad &j=1,\ldots,N_{\rho}.
 \end{align}

 Associated with these grids are  quadratures with weight functions $w(\rho)=\rho$  and $w(F_\rho)=F_\rho$ in direct and Fourier space, respectively, with the corresponding weights
 \begin{align}
 w_j^{(\rho)} &= \frac{2R^2}{\alpha_{N_\rho+1}^2}\frac{1}{J_1(\alpha_j)^2},\\
 w_j^{(F_\rho)}&= \frac{2}{R^2}\frac{1}{J_1(\alpha_j)^2},
 \end{align}
 see  \cite{Lemoine1994a,Nygaard2004a}.
 
The in-plane (i.e.~two dimensional)  Fourier transformation of a radially symmetric function, $f(\rho)$, is given by
\begin{align}
\tilde{f}(F_\rho)&=\int_0^\infty d\rho\,\rho \int_0^{2\pi} d\phi\, e^{iF_\rho\rho\cos\phi}f(\rho),\\
&=2\pi\left[\int_0^\infty d\rho\,\rho J_0(F_\rho\rho)f(\rho)\right],\label{Hanktransf}
\end{align} 
where $\phi$ is the in-plane angular coordinate.  The term in the square brackets in Eq.~(\ref{Hanktransf}) is a Hankel transformation. Making use of the quadrature (\ref{Hanktransf}) is approximately given by
\begin{equation}
\tilde{f}(F_{\rho,i})\approx\sum_jH_{ij}f(\rho_j),
\end{equation}
 where the transformation matrix is  
\begin{equation}
H_{ij} =\frac{4\pi R^2}{\alpha_{N_\rho+1}^2}\frac{J_0\left(\alpha_i\alpha_j/\alpha_{N_\rho+1}\right)}{J_1(\alpha_j)^2}.
\end{equation}
The inverse transformation, i.e.~$\tilde{f}(F_{\rho,i})\to f(\rho_j)$, is given by the matrix
\begin{equation}
\left(H^{-1}\right)_{ij} =\frac{1}{4\pi R^2}\frac{J_0\left(\alpha_i\alpha_j/\alpha_{N_\rho+1}\right)}{J_1(\alpha_j)^2}.
\end{equation}

For the case of the radial momentum variable $k_\rho$, the grid is similarly given by
\begin{align}
 k_{\rho_j}&=\frac{\alpha_j K}{\alpha_{N_{k_\rho}+1}},\quad &j=1,\ldots,N_{k_\rho},
\end{align}
 where $K$ is the range of the momentum grid.
The above discussion regarding the quadrature of quantities on the position grid immediately applies to the momentum case with the replacement $N_\rho\to N_{k_\rho}$ and $R\to K$. For future reference, we denote  $w_i^{(k_\rho)} $ as the quadrature weights.

\subsection{Evaluating the direct term}
Here we demonstrate how (\ref{PhiD}) is evaluated numerically. First, we obtain the density,
\begin{align}
 n(\rho,z) &=\int_{0}^{\infty} d{k_z}\int_0^\infty dk_\rho\frac{k_\rho}{2\pi^2}W(\rho,z,k_\rho,k_z),\\
 \Rightarrow n_{qr} &=\frac{\Delta_{{k_z}}}{2\pi^2}\sum_{st}w_s^{(k_\rho)}W_{qrst},
\end{align}
where we have made use of the momentum variable quadratures to perform the integrations.

Then we note that the convolution is given by Fourier transforms as $\Phi_D(\bx) = \mathcal{F}^{-1}[\tilde{U}_{dd}(\k)n(\k)]$. Ronen et al. \cite{Ronen2006a} have shown that the discontinuity of $\tilde{U}_{dd}(\k)$ at the origin results in slow convergence without a cutoff in the real space interaction potential. Using a cutoff, $L$, that is larger than the system has no physical consequences, and results in the corrected Fourier transformed interaction:
\begin{align}
    \tilde{U}^{L}_{dd}(\k)=\frac{1}{3}C_{dd}\left[1+3\frac{\cos(Lk)}{(Lk)^2} - 3 \frac{\sin(Lk)}{(Lk)^3} \right](3\cos\theta^2_\k-1),
\end{align}

Then we have that (\ref{PhiD}) is given by
\begin{align}
\Phi_{D,mn}\equiv&\Phi_D(\rho_m,z_n),\\
=&\sum_{ij}\left(H^{-1}\right)_{mi}\left(T^{-1}\right)_{nj}\tilde{U}^L_{dd}(F_{\rho,i},F_{z,j}) \nonumber\\
&\times\sum_{qr}H_{iq}T_{jr}n_{qr}.\label{e:PhiDmn}
\end{align} 

Our calculations for anisotropic traps require the cutoff to be larger than any of the dimensions of our cloud. As our aspect ratios are up to a factor of ten, it is computationally efficient to avoid evaluation of the terms in \eqref{e:PhiDmn} where the density is known to be negligibly small.

\subsection{Evaluating the exchange term}
We now consider (\ref{PhiE}). Unlike the direct term, which acts on operate on a marginal (i.e.~the density), the exchange interaction acts on the complete Wigner distribution. We adopt the same notation for the direction term, i.e.
\begin{equation}
\Phi_{E,qrmn}\equiv\Phi_E(\rho_q,z_r,k_{\rho_m},k_{z_n}).
\end{equation}
We consider an analytic simplification of the exchange term
%\begin{widetext}
\begin{align}
\Phi_E(\x,\k)&=\int \frac{ d\kd}{(2\pi)^3}W(\x,\kd)\tilde{U}_{dd}(\k-\kd), \\
&=\frac{C_{dd}}{3}\int  \frac{ d\kd}{(2\pi)^3}W(\x,\kd)\left[\frac{3(k_z-k_z')^2}{|\k-\kd|^2}-1\right],\\
%&=\frac{C_{dd}}{3(2\pi)^3}\int_{-\infty}^{\infty}dk_z'\int_0^{\infty}k_{\rho}'dk_{\rho}'\int_0^{2\pi}dk_{\phi}'W(\x,\kd)\left[\frac{3(k_z-k'_z)^2}{k^2+k^{\prime2}-2k_\rho k_\rho^\prime\cos(k_\phi-k_\phi^\prime)-2k_zk_z^\prime}-1\right],\\
&=\int_{0}^{\infty}dk_z'\int_0^{\infty}k_{\rho}'dk_{\rho}' W(\x,k_\rho^\prime,k_z^\prime)\times\notag\\&\left[\Lambda(k_\rho,k_z,k_\rho^\prime,k_z^\prime)+\Lambda(k_\rho,k_z,k_\rho^\prime,-k_z^\prime)\right],\label{diPhiE}
\end{align}
where we have integrated out $k'_\phi$ and used that
$W(\x,\k)$ is even in $k_z$ and has no $k_\phi$ dependence, and defined
\begin{align}
&\Lambda(k_\rho,k_z,k_\rho^\prime,k_z^\prime)\equiv\frac{C_{dd}}{3(2\pi)^2}\times\notag\\
&\hspace{1cm}\left[\frac{3(k_z-k'_z)^2}{\sqrt{(k^2+k^{\prime2}-2k_zk_z^\prime)^2-4k_\rho^2k_\rho^{\prime2}}}-1\right].
\end{align}
We note that Eq.~(\ref{diPhiE}) is consistent with the functional derivative  of the expression Zhang and Yi have developed for the exchange energy (see Sec.~IV of \cite{Zhang2009a}).

Numerically the exchange term is  evaluated as
\begin{equation}
\Phi_{E,qrmn}= \Delta_{k_z}\sum_{st} w_s^{(k_\rho)}W_{qrst}\Lambda_{mnst},
\end{equation}
where $\Lambda_{mnst}=\Lambda(k_{\rho_m},k_{z_n},k_{\rho_s},k_{z_t})$.

\subsection{Choice of number of grid points}
Using the gaussian ansatz of \cite{Endo2010a}, we compared the total direct and exchange energies to the exact result at $T=0.5T_F^0$. We found that the relative error in the direct energy decreased exponentially with  $\{N_\rho,N_z\}$ and was less than $10^{-12}$ with $\{N_\rho,N_z\}=\{24,24\}$. The relative error in the exchange energy decreased algebraically with $\{N_{k_\rho},N_{k_z}\}$, $N_{k_\rho}=0.6N_{k_z}$ was optimum and with $\{N_{k_\rho},N_{k_z}\} = \{48,80\}$ the relative error was up to $10^{-4}$.  

For the results presented in Section \ref{s:results}, the Wigner function and dispersion relation are calculated on four dimensional grids with sizes of $\{N_\rho,N_z,N_{k_\rho},N_{k_z}\}=\{40, 40, 48, 80\}$.  For each temperature these calculations can take approximately 10 hours to converge to self-consistency on an eight-core 2.8 GHz Xeon processor.

\bibliographystyle{apsrev4-1}

\end{document}